\documentclass{kluwer}

\begin{document}      

\begin{article}                                                     
\begin{opening}         
\title{SOME BIANCHI TYPE I VISCOUS FLUID COSMOLOGICAL MODELS WITH A 
VARIABLE COSMOLOGICAL CONSTANT}
\author{ANIRUDH PRADHAN\thanks{Corresponding Author}}
\runningtitle{SOME BIANCHI TYPE I VISCOUS FLUID COSMOLOGICAL MODELS ....}
\runningauthor{A. PRADHAN AND P. PANDEY }
\author{PURNIMA PANDEY}
\institute{Department of Mathematics, Hindu Post-graduate College,
Zamania-232 331, Ghazipur, U. P., India; \\
E-mail:acpradhan@yahoo.com, pradhan@iucaa.ernet.in}


\begin{abstract} 
Some Bianchi type I viscous fluid cosmological models with a variable cosmological 
constant are investigated in which the expansion is considered only in two direction
i.e. one of the Hubble parameter $(H_{1} = \frac{A_{4}}{A})$ is zero. The viscosity
coefficient of bulk viscous fluid is assumed to be a power function of mass density 
whereas the coefficient of shear viscosity is considered as constant in first 
case whereas in other case it is taken as proportional to scale of expansion in the
model. The cosmological constant $\Lambda$ is found to be positive and is a decreasing 
function of time which is supported by results from recent supernovae Ia observations. 
Some physical and geometric properties of the models are also discussed.
\end{abstract}
\end{opening}

\section{Introduction}
\vspace*{-0.5pt}
\noindent
Models with a dynamic cosmological term $\Lambda (t)$ are becoming popular as 
they solve the cosmological constant problem in a natural way. There are significant 
observational evidence for the detection of Einstein's cosmological constant, 
$\Lambda$ or a component of material content of the universe that varies slowly with 
time and space to act like $\Lambda$. Recent cosmological observations by High -Z 
Supernova Team and Supernova Cosmological Project (Garnavich {\it et al.}, 1998; 
Perlmutter {\it et al.}, 1997, 1998, 1999; Riess {\it et al.}, 1998; Schmidt {\it et al.}, 
1998) suggest the existence of a positive cosmological constant $\Lambda$ with 
magnitude $\Lambda (G \hbar /c^{3}) \approx 10^{-123}$. These observations on magnitude 
and red-shift of type Ia supernova suggest that our universe may be an accelerating  
function of the cosmological density in the form of the cosmological $\Lambda$-term. 
Earlier researches on this topic are contained in Zeldovich (1968), Weinberg (1972), 
Dolgov (1983, 1990), Bertolami (1986), Ratra and Peebles (1988), Carroll, Press and 
Turner (1992). Some of the recent discussions on the cosmological constant ``problem'' 
and consequences on cosmology with a time-varying cosmological constant have been discussed 
by Dolgov (1993,1997), Tsagas and Maartens (2000), Sahni and Starobinsky (2000), Peebles 
(2003), Padmanabhan (2003), Vishwakarma (1999, 2000, 2001, 2002), and Pradhan {\it et al.} 
(2001, 2002, 2003, 2004). This motivates us to study the cosmological models in which
 $\Lambda$ 
varies with time. \\ 

The distribution of matter can be satisfactorily described by a perfect fluid
due to the large scale distribution of galaxies in our universe. However, 
observed physical phenomena such as the large entropy per baryon and the 
remarkable degree of isotropy of the cosmic microwave background radiation, 
suggest analysis of dissipative effects in cosmology. Furthermore, there are
several processes which are expected to give rise to viscous effects. These
are the decoupling of neutrinos during the radiation era and the decoupling
of radiation and matter during the recombination era. Bulk viscosity is
associated with the GUT phase transition and string creation. Misner (1967, 1968)
has studied the effect of viscosity on the evolution of cosmological models. The
role of viscosity in cosmology has been investigated by Weinberg (1971). Nightingale
(1973), Heller and Klimek (1975) have obtained a viscous universes without initial
singularity. The model studied by Murphy (1973) possessed an interesting feature in 
which  big bang type of singularity of infinite space-time curvature does not occur 
to be a finite past. However, the relationship assumed by Murphy between the
viscosity coefficient and the matter density is not acceptable at large 
density. Roy and Prakash (1977) have investigated a plane symmetric cosmological models 
representing a viscous fluid with free gravitational field of non-degenerate Petrov 
type I in which coefficient of shear viscosity is constant. Bali and Jain (1987, 1988)
have obtained some expanding and shearing viscous fluid cosmological models in which 
coefficient of shear viscosity is proportional to rate of expansion in the model and
the free gravitational field is Petrov type D and non-degenerate. The effect of bulk 
viscosity on the cosmological evolution has been investigated by a number of authors 
in the framework of general theory of relativity (Padmanabhan and Chitre, 1987; Johri 
and Sudarshan, 1988; Maartens, 1995; Zimdahl, 1996; Pradhan {\it et al.}, 1997; 
Kalyani and Singh, 1997; Singh {\it et al.}, 1998; Pradhan {\it et al.}, 2001, 2002). 
This motivates to study cosmological bulk viscous fluid model.\\

Recently Bali and Jain (1997) have investigated Bianchi type I viscous fluid cosmological
models. Motivated by the situations discussed above, in this paper, we shall focus
on the problem with varying cosmological constant in presence of bulk and
shear viscous fluid in an expanding universe. We do this by extending the work
of Bali and Jain (1997) by including varying cosmological constant and the coefficient
of bulk viscosity as function of time. This paper is organized as follows: The
metric and the field equations are presented in section 2. In section 3, we
deal with the solution of the field equations in presence of viscous fluid.
In section $3.1$, we consider the coefficient of shear viscosity ($\eta$) as
constant whereas in section $3.2$, $\eta$ is taken as proportional to scale of 
expansion in the model. Section 4 includes the solution in absence of shear viscosity. 
In section 5, we have given the concluding remarks.    \\   
\section{The Metric and Field  Equations}
We consider the Bianchi type I metric in the form
\begin{equation}
\label{eq1}
ds^{2} = - dt^{2} + dx^{2} + B^{2}dy^{2} + C^{2}dz^{2},
\end{equation}
where $B$ and $C$ are functions of $t$ alone. This metric depicts the case
when one of the Hubble parameters (here $H_{1} = \frac{A_{4}}{A}$) is zero,
i. e. the expansion is only in two directions. The kinematic parameters are
then related as $\theta = - 3\sigma^{1}_{1}$. This condition leads to the metric 
(\ref{eq1}). \\
The Einstein's field equations (in gravitational units $c = 1$, $G = 1$) read as
\begin{equation}
\label{eq2}
R^{j}_{i} - \frac{1}{2} R g^{j}_{i} + \Lambda g^{j}_{i} = - 8\pi T^{j}_{i},
\end{equation}
where $R^{j}_{i}$ is the Ricci tensor; $R$ = $g^{ij} R_{ij}$ is the
Ricci scalar; and $T^{j}_{i}$ is the stress energy-tensor in the presence
of bulk stress given by Landau and Lifshitz (1963) 
\[
T^{j}_{i} = (\rho + p)v_{i}v^{j} + p g^{j}_{i} - \eta \left ({v^{j}_{i ;}} + {v^j}_{;i} 
+ v^{j} v^{l} v_{i;l} + v_{i} v^{l} {v^j}_{;l}\right) 
\]
\begin{equation}
\label{eq3}
- \left (\xi - \frac{2}{3} \eta \right)
 ~ \theta ({g^j}_{i} + v_{i} v^{j}).
\end{equation}
Here $\rho$, $p$, $\eta$ and $\xi$ are the energy density,
isotropic pressure, coefficient of shear viscosity and bulk viscous 
coefficient respectively and $v_{i}$ is the flow vector satisfying 
the relations
\begin{equation}
\label{eq4}
g_{ij} v^{i}v^{j} = - 1.
\end{equation}
The semicolon $(;)$ indicates covariant differentiation. We choose the coordinates
to be comoving, so that 
\begin{equation}
\label{eq5}
v^1 = 0 = v^2 = v^3, v^4 = 1.
\end{equation}
The Einstein's field equations (\ref{eq2}) for the line element (\ref{eq1})
has been set up as
\begin{equation}
\label{eq6}
- 8 \pi \left[p  - \left (\xi - \frac{2}{3}\eta \right) \theta\right] =
\frac{B_{44}}{B} + \frac{C_{44}}{C} + \frac{B_{4}C_{4}}{BC} + \Lambda, 
\end{equation}
\begin{equation}
\label{eq7}
- 8 \pi \left[p - 2\eta \frac{B_{4}}{B} - \left (\xi - \frac{2}{3}\eta \right) 
\theta\right] = \frac{C_{44}}{C} + \Lambda,
\end{equation}
\begin{equation}
\label{eq8}
- 8 \pi \left[p - 2\eta \frac{C_{4}}{C} - \left (\xi - \frac{2}{3}\eta \right) 
\theta\right] = \frac{B_{44}}{B} + \Lambda,
\end{equation}
\begin{equation}
\label{eq9}
8 \pi \rho = \frac{B_{4} C_{4}}{B C} + \Lambda, 
\end{equation}
where the suffix $4$ at the symbols $A$ and $B$ denotes ordinary 
differentiation with respect to $t$ and $\theta$ is the scalar of expansion given by
\begin{equation}
\label{eq10}
\theta = {v^{i}}_{;i}.
\end{equation}
\section{Solution of the Field Equations}
We have revisited the solutions obtained by Bali and Jain (1997).
Equations (\ref{eq6}) - (\ref{eq9}) are four independent equations in 
seven unknowns $A$, $B$, $\rho$, $p$, $ \eta$, $\xi$ and $\Lambda$. For 
the complete determinacy of the system, we need three extra conditions. 
The research on exact solutions is based on some physically reasonable 
restrictions used to simplify the Einstein equations. \\
From Eqs. (\ref{eq6}) - (\ref{eq8}), we obtain
\begin{equation}
\label{eq11}
\frac{B_{44}}{B} + \frac{B_{4}C_{4}}{BC} = - 16 \pi \eta \frac{B_{4}}{B}, 
\end{equation}
and 
\begin{equation}
\label{eq12}
\frac{B_{44}}{B} - \frac{C_{44}}{C} = -16 \pi \eta \left(\frac{B_{4}}{B} - 
\frac{C_{4}}{C}\right).
\end{equation}
Equation (\ref{eq12}) on integration leads to
\begin{equation}
\label{eq13}
C^{2}\left(\frac{B}{C}\right)_{4} = L e^{-16\pi\eta t},
\end{equation}
where $L$ is a constant of integration. Setting $BC = \mu$ and $\frac{B}{C} 
= \nu$ in Eqs. (\ref{eq11}), (\ref{eq12}) and (\ref{eq13}) lead to
\begin{equation}
\label{eq14}
\mu_{44} + 16 \pi \eta \mu_{4} = 0,
\end{equation}
and
\begin{equation}
\label{eq15}
\frac{\nu_{4}}{\nu} = \frac{L}{\mu} e^{-16 \pi \eta t}. 
\end{equation}
Equation (\ref{eq14}) on integration leads to
\begin{equation}
\label{eq16}
\mu_{4} = M e^{-16\pi \eta t},
\end{equation}
where $M$ is an integrating constant. \\
Here we consider two cases: \\
\subsection{CASE I : LET $\eta$ = CONSTANT = $a$ (SAY)}
In this case Eq. (\ref{eq16}) on integration gives
\begin{equation}
\label{eq17}
\mu = N - \frac{M}{16\pi a} e^{-16 \pi at}, 
\end{equation}
where $N$ is a constant of integration. \\
Equation (\ref{eq15}) on integration leads to
\begin{equation}
\label{eq18}
\nu = \alpha(16\pi aN - M e^{-16\pi at})^{\frac{L}{M}},
\end{equation}
where $\alpha$ is an integrating constant. \\
Now we set
\begin{equation}
\label{eq19}
e^{-16\pi at} = \cos2\sqrt{16\pi a}\tau,
\end{equation}
\begin{equation}
\label{eq20}
\alpha = \frac{1}{(16\pi a)^{\frac{L}{M}}},
\end{equation}
and
\begin{equation}
\label{eq21} 
N = \frac{M}{16\pi a}.
\end{equation}
Using the above Eqs. (\ref{eq19}) - (\ref{eq21}) in (\ref{eq17}) and 
(\ref{eq18}), we obtain
\begin{equation}
\label{eq22}
\mu = 2M\left(\frac{\sin\sqrt{16\pi a}\tau}{\sqrt{16\pi a}}\right)^{2},
\end{equation}
and
\begin{equation}
\label{eq23}
\nu = (2M)^{\frac{L}{M}}\left(\frac{\sin\sqrt{16\pi a}\tau}{\sqrt{16\pi a}}\right)^
{\frac{2L}{M}}.
\end{equation}
From above Eqs. (\ref{eq22}) and (\ref{eq23}), we obtain
\begin{equation}
\label{eq24}
B^{2} = \mu \nu = (2M)^{1 + \frac{L}{M}}\left(\frac{\sin \sqrt{16\pi a}\tau}
{\sqrt{16\pi a}}\right)^{2 + \frac{2L}{M}}
\end{equation}
\begin{equation}
\label{eq25}
C^{2} = \frac{\mu}{\nu} = (2M)^{1 - \frac{L}{M}}\left(\frac{\sin \sqrt{16\pi a}\tau}
{\sqrt{16\pi a}}\right)^{2 - \frac{2L}{M}}
\end{equation}
Hence the metric (\ref{eq1}) reduces to the form
\[
ds^{2} = - 4\left(\frac{\tan2\sqrt{16\pi a}\tau}{\sqrt{16\pi a}}\right) d\tau^{2} 
+ dx^{2} + (2M)^{1 + \frac{L}{M}}\left(\frac{\sin\sqrt{16\pi a}\tau}
{\sqrt{16\pi a}}\right)^{2 + \frac{2L}{M}} dy^{2}
\]
\begin{equation}
\label{eq26}
+ (2M)^{1 - \frac{L}{M}}\left(\frac{\sin\sqrt{16\pi a}\tau}
{\sqrt{16\pi a}}\right)^{2 - \frac{2L}{M}} dz^{2}
\end{equation}
After suitable transformation of coordinates metric (\ref{eq26}) reduces to the form
\begin{equation}
\label{eq27}
ds^{2} = - 4\left(\frac{\tan 2kT}{k}\right) dT^{2} + dX^{2} + \left(\frac{\sin kT}{k}
\right)^{2 + \frac{2L}{M}} dY^{2} + \left(\frac{\sin kT}{k}\right)^{2 - \frac{2L}{M}} 
dz^{2},
\end{equation}
where $k = \sqrt{16\pi a}$. \\
The pressure and density for the model (\ref{eq27}) are given by
\[
8\pi p = \frac{1}{4M^{2}}\left(\frac{k}{\sin kT}\right)^{4}\biggl[M(M - L) 
\sin^{2}kT \cos 2kT - \frac{(L + M)}{4}\cos 2kT \times
\]
\begin{equation}
\label{eq28}
\{(L + M)\cos 2kT - 2M\} + 16 \pi M^{2}\left(\xi - \frac{2}{3}a\right)
\cos 2kT \left(\frac{\sin kT}{k}\right)^{2}\biggr] - \Lambda
\end{equation}
\begin{equation}
\label{eq29}
8\pi \rho = \frac{1}{4M^{2}}\left(\frac{k}{\sin kT}\right)^{4}\left[\frac{(M^{2} 
- L^{2})}{4}\cos^{2} 2kT\right] + \Lambda
\end{equation}
For the specification of $\xi$, we assume that the fluid obeys an equation of 
state of the form
\begin{equation}
\label{eq30}
p = \gamma \rho,
\end{equation}
where $\gamma (0 \leq \gamma \leq 1)$ is constant. Thus, given $\xi(t)$ we can 
solve for the cosmological parameters. In most of the investigation involving 
bulk viscosity is assumed to be a simple power function of the energy density 
(Pavon, 1991; Maartens, 1995; Zimdahl, 1996) 
\begin{equation}
\label{eq31}
\xi(t) = \xi_{0} \rho^{n},
\end{equation}
where $\xi_{0}$ and $n$ are constants. If $n = 1$, Equation (\ref{eq23}) may 
correspond to a radiative fluid (Weinberg, 1972). However, more realistic models 
(Santos, 1985) are based on $n$ lying in the regime $0 \leq n \leq \frac{1}{2}$. \\
On using (\ref{eq31}) in (\ref{eq28}), we obtain
\[
8\pi p = \frac{1}{4M^{2}}\left(\frac{k}{\sin kT}\right)^{4}\biggl[M(M - L) 
\sin^{2}kT \cos 2kT - \frac{(L + M)}{4}\cos 2kT \times 
\]
\begin{equation}
\label{eq32}
\{(L + M)\cos 2kT - 2M \} + 16 \pi M^{2}\left(\xi_{0}\rho^{n} - 
\frac{2}{3}a\right)\cos 2kT \left(\frac{\sin kT}{k}\right)^{2}\biggr] - \Lambda
\end{equation}
\subsubsection{Model I: Solution for $\xi = \xi_{0}$}
When $n = 0$, Equation (\ref{eq31}) reduces to $\xi = \xi_{0}$ = constant. 
Hence in this case Equation (\ref{eq32}), with the use of (\ref{eq29}) and 
(\ref{eq30}), leads to
\[
64 \pi M^{2}(1 + \gamma)\rho = \left(\frac{k}{\sin kT}\right)^{4} \cos 2kT
\biggl[M^{2} - L^{2} + 2(L^{2} + M^{2})\sin^{2}kT + 
\]
\begin{equation}
\label{eq33}
32\pi M^{2}\left(\xi_{0} - \frac{2a}{3}\right)\left(\frac{\sin kT}{k}\right)^{2}\biggr] 
\end{equation}
Eliminating $\rho(t)$ between Eqs. (\ref{eq29}) and (\ref{eq33}), we have
\[
2 (1 + \gamma)\Lambda = \cos 2kT \biggl[(1 - \gamma)(M^{2} - L^{2}) + 
\]
\begin{equation}
\label{eq34}
2\{(3M^{2}+ L^{2}) + (M^{2} - L^{2})\gamma\}\sin^{2}kT + 64\pi M^{2}\left(\xi_{0} - 
\frac{2a}{3}\right)\left(\frac{\sin kT}{k}\right)^{2}\biggr] 
\end{equation}
\subsubsection{Model I: Solution for $\xi = \xi_{0}\rho$}
When $n = 1$, Equation (\ref{eq31}) reduces to $\xi = \xi_{0}\rho$. 
Hence in this case Equation (\ref{eq32}), with the use of (\ref{eq29}) and 
(\ref{eq30}), leads to
\[
16\pi \rho = \frac{k^{4}\cos2kT}{M^{2}[2(1 + \gamma)\sin^{2}kT - k^{2}\xi_{0} 
\cos 2kT] \sin^{2}kT} \times  
\]
\begin{equation}
\label{eq35}
\left[\frac{M^{2} - L^{2}}{2} + (L^{2} + M^{2})\sin^{2}kT - \frac{8\pi a}{3} 
\cos 2kT \left(\frac{k}{\sin kT}\right)^{2}\right]
\end{equation}
Eliminating $\rho(t)$ between (\ref{eq29}) and (\ref{eq35}), we have
\[
\Lambda = - \frac{(M^{2} - L^{2})\cos^{2}2kT}{16 M^{2}} \left(\frac{k}{\sin kT} 
\right)^{4} + 
\]
\[
\frac{k^{4}\cos2kT}{2M^{2}[2(1 + \gamma)\sin^{2}kT - k^{2}\xi_{0} 
\cos 2kT] \sin^{2}kT} \times  
\]
\begin{equation}
\label{eq36}
\left[\frac{M^{2} - L^{2}}{2} + (L^{2} + M^{2})\sin^{2}kT - \frac{8\pi a}{3} 
\cos 2kT \left(\frac{k}{\sin kT}\right)^{2}\right]
\end{equation}
Our study for these models shows constant value of cosmological constant ($\Lambda$)
for large values of time and do not decrease with time (this means that the universe
is not expanding or may be steady state condition). In this case, detailed study shows
that the scalar of expansion $\theta$ does not increase with time. Our study is 
inconsistent with work done by Bali and Jain (1997). Bali and Jain claim that the
universe is expanding which does not match with our result. Also the claim of minimum
and maximum expansion rate in $\theta$ is reflection of periodicity of trigonometrical 
functions involved there. We are trying to find feasible interpretation and situations 
relevant to this case. Further study is in progress. \\ 
{\bf Some Physical Aspects of the Models}: \\
With regard to the kinematic properties of the velocity vector $v^{i}$ in the metric 
(\ref{eq27}), a straight forward calculation leads to the following expressions for the
scalar of expansion $(\theta)$ and for the shear $(\sigma)$ of the fluid.
\begin{equation}
\label{eq37}
\theta = \frac{1}{2} \cos 2kT \left(\frac{k}{\sin kT}\right)^{2}
\end{equation}
\begin{equation}
\label{eq38}
\sigma^{2} = \frac{1}{48}\left(1 + \frac{3L^{2}}{M^{2}}\right) \cos^{2} 2kT \left(\frac{k}
{\sin kT}\right)^{4}
\end{equation}
The rotation $\omega$ is identically zero. The non-vanishing components of conformal
curvature tensor are
\begin{equation}
\label{eq39}
C^{12}_{12} = \frac{\cos 2kT}{48 M^{2}} \left(\frac{k}{\sin kT}\right)^{4}\left[M^{2} 
- L^{2} - 2L(3M - L)\sin^{2} kT\right]
\end{equation}
\begin{equation}
\label{eq40}
C^{13}_{13} = \frac{\cos 2kT}{48 M^{2}} \left(\frac{k}{\sin kT}\right)^{4}\left[M^{2} 
- L^{2} + 2L(3M + L)\sin^{2} kT\right]
\end{equation}
\begin{equation}
\label{eq41}
C^{14}_{14} = \frac{\cos 2kT}{24 M^{2}} \left(\frac{k}{\sin kT}\right)^{4}\left[L^{2} 
- M^{2} - 2L^{2}\sin^{2} kT\right]
\end{equation}
The models represent shearing, non-rotating and Petrov type I non-degenerate in
general, in which the flow is geodetic. However, if $L = 0$ then spacetime reduces 
to Petrov type $D$. The model starts expanding at $T > 0$ but the initial expansion 
is slow. When $T$ is closer to $\pi/2k$, it has stiff rise in the expansion and then 
decreases. This shows the case of oscillation. It is observed that the expansion is 
minimum at $T = 0$ or $T = \pi/k$. The large values of $\theta$ near
$T = \pi/2k$ is reflection of trigonometric property. But expansion remains finite.
The expression for $\frac{\sigma}{\theta}$ and $\frac{\rho}{\theta^{2}}$ are found 
to be
\begin{equation}
\label{eq42}
\frac{\sigma}{\theta} = \frac{1}{2\sqrt{3}} \left(1 + \frac{3L^{2}}{M^{2}}\right)^
{\frac{1}{2}}
\end{equation}
\begin{equation}
\label{eq43}
\frac{\rho}{\theta^{2}} = \frac{1}{32 \pi M^{2} \cos^{2} 2kT} \left[(M^{2} - L^{2})
\cos^{2} 2kT 
+ 16 M^{2} \Lambda \left(\frac{\sin kT}{k}\right)^{4}\right] 
\end{equation}
From Equation (\ref{eq42}), it can be seen that shear is proportional to scalar 
of expansion $\theta$ in the model. Since $\lim_{T \rightarrow \infty} \frac{\sigma}
{\theta} \neq 0$, hence the models do not approach isotropy for large values of $T$.  \\
The rate of expansion $H_{i}$ (Hubble parameters) in the direction of $X$, $Y$, $Z$ 
are given by
\begin{equation}
\label{eq44}
H_{1} = 0
\end{equation}
\begin{equation}
\label{eq45}
H_{2} = \frac{(M + L)}{4M T^{2}}
\end{equation}
\begin{equation}
\label{eq46}
H_{2} = \frac{(M - L)}{4M T^{2}}
\end{equation}
\subsection{CASE II: LET $\eta = \beta \theta$, WHERE $\beta$ IS AN ARBITRARY CONSTANT.}
In this case Equation (\ref{eq14}) reduces to
\begin{equation}
\label{eq47}
\mu \mu_{44} + 16 \pi \beta \mu^{2}_{4} = 0, 
\end{equation}
which on integration leads to
\begin{equation}
\label{eq48}
\mu = [(1 + 16\pi \beta)(k_{1}t + k_{2})]^{\frac{1}{(1 + 16\pi \beta)}},
\end{equation}
where $k_{1}$ and $k_{2}$ are constants of integration. \\
Equation (\ref{eq12}) reduces to
\begin{equation}
\label{eq49}
\mu \left(\frac{\nu_{4}}{\nu}\right)_{4} = - (1 + 16\pi \beta)\frac{\mu_{4}}{\mu},
\end{equation}
which on integration leads to
\begin{equation}
\label{eq50}
\nu = k_{3}(k_{1}t + k_{2})^{\frac{k_{4}}{k_{1}(1 + 16\pi \beta)}},
\end{equation}
where $k_{3}$ and $k_{4}$ are constants of integration. \\
From Equations (\ref{eq48}) and (\ref{eq50}), we obtain
\begin{equation}
\label{eq51}
B^{2} = \mu \nu = k_{3} k_{5} (k_{1} t + k_{2})^{k_{6}(k_{1} + k_{4})},
\end{equation}
\begin{equation}
\label{eq52}
C^{2} = \frac{\mu}{\nu} = \frac{k_{5}}{k_{3}} (k_{1} t + k_{2})^{k_{6}(k_{1} - k_{4})},
\end{equation}
where
\begin{equation}
\label{eq53}
k_{5} = (1 + 16\pi \beta)^{\frac{1}{(1 + 16\pi \beta)}},
\end{equation}
\begin{equation}
\label{eq54}
k_{6} = \frac{1}{k_{1}(1 + 16\pi \beta)}.
\end{equation}
Hence the metric (\ref{eq1}) reduces to
\begin{equation}
\label{eq55}
ds^{2} = - dt^{2} + dx^{2} + k_{3}k_{5}(k_{1}t + k_{2})^{k_{6}(k_{1} + k_{4})} dy^{2}
+ \frac{k_{5}}{k_{3}} (k_{1} t + k_{2})^{k_{6}(k_{1} - k_{4})} dz^{2}
\end{equation}
After suitable transformation of coordinates, the metric (\ref{eq55}) takes the form
\begin{equation}
\label{eq56}
ds^{2} = - \frac{1}{k^{2}_{1}} dT^{2} + dX^{2} + T^{k_{6}(k_{1} + k_{4})} dY^{2}
+ T^{k_{6}(k_{1} - k_{4})} dZ^{2}
\end{equation}
The pressure and density for the model (\ref{eq56}) are given by 
\[
8\pi p = - \frac{k^{2}_{1}k_{6}(k_{1} + k_{4})}{4 T^{2}}(k_{1}k_{6} + k_{4}k_{6} - 2)
\]
\begin{equation}
\label{eq57}
+ \frac{8\pi k^{2}_{1}k_{6}}{T}\biggl[\frac{\beta k_{1}k_{6}}{3T}(k_{1} - k_{4})
+ \xi \biggr] - \frac{k_{2}}{2k^{2}_{5} T^{2k_{1} k_{6}}} - \Lambda
\end{equation}
\begin{equation}
\label{eq58}
8 \pi \rho = \frac{k^{2}_{1}k^{2}_{6}(k^{2}_{1} - k^{2}_{4})}{4 T^{2}} + \Lambda
\end{equation}
On using (\ref{eq31}) in (\ref{eq57}), we obtain
\[
8\pi p = - \frac{k^{2}_{1}k_{6}(k_{1} + k_{4})}{4 T^{2}}(k_{1}k_{6} + k_{4}k_{6} - 2)
\]
\begin{equation}
\label{eq59}
+ \frac{8\pi k^{2}_{1}k_{6}}{T}\biggl[\frac{\beta k_{1}k_{6}}{3T}(k_{1} - k_{4})
+ \xi_{0}\rho^{n} \biggr] - \frac{k_{2}}{2k^{2}_{5} T^{2k_{1} k_{6}}} - \Lambda
\end{equation}
\subsubsection{Model I: Solution for $\xi = \xi_{0}$}
When $n = 0$, Equation (\ref{eq59}), with the use of (\ref{eq58}) and 
(\ref{eq30}), leads to
\[
8\pi (1 + \gamma)\rho = \frac{k^{2}_{1}k_{6}(k_{1} + k_{4})
(1 - k_{4}k_{6})}{2T^{2}} 
\]
\begin{equation}
\label{eq60}
+ \frac{8\pi k^{2}_{1}k_{6}}{T}\biggl[\frac{\beta k_{1}k_{6}}{3T}(k_{1} - k_{4})
+ \xi_{0} \biggr] - \frac{k_{2}}{2k^{2}_{5} T^{2k_{1} k_{6}}}
\end{equation}
Eliminating $\rho(t)$ between Equations (\ref{eq58}) and (\ref{eq60}), we have
\[
(1 + \gamma)\Lambda = \frac{k^{3}_{1}k_{6}(k_{1} + k_{4})}{T} 
\biggl[2 - \{k_{1} + k_{4} + (k_{1} - k_{4})\gamma\}k_{6}\biggr] 
\] 
\begin{equation}
\label{eq61}
+ \frac{8\pi k^{2}_{1}k_{6}}{T}\biggl[\frac{\beta k_{1}k_{6}}{3T}(k_{1} - k_{4})
+ \xi_{0} \biggr] - \frac{k_{2}}{2k^{2}_{5} T^{2k_{1} k_{6}}}
\end{equation}
\subsubsection{Model I: Solution for $\xi = \xi_{0}\rho$}
When $n = 1$, Equation (\ref{eq59}), with the use of (\ref{eq58}) and 
(\ref{eq30}), leads to
\[
8\pi\left[1 + \gamma - \frac{k^{2}_{1}k_{6} \xi_{0}}{T}\right] \rho =
\frac{k^{2}_{1}k^{2}_{6}}{6T^{2}} \biggl[3(k_{1} + k_{4})(1 - k_{4}k_{6})
\]
\begin{equation}
\label{eq62}
+  16\pi \beta k_{1}(k_{1} - 3k_{4})\biggr] - \frac{k_{2}}{2k^{2}_{5} 
T^{2k_{1} k_{6}}}
\end{equation}
Eliminating $\rho(t)$ between (\ref{eq58}) and (\ref{eq62}), we have
\[
\biggl[1 + \gamma - \frac{k^{2}_{1}k_{6} \xi_{0}}{T}\biggr] \Lambda =
\frac{k^{2}_{1}k_{6}(k_{1} + k_{4})}{4T^{2}} \biggl[2 - (k_{1} + k_{4})k_{6} 
\]
\begin{equation}
\label{eq63}
- (k_{1} - k_{4})k_{6}\{\gamma - \frac{k^{2}_{1} k_{6}\xi_{0}}{T}\}\biggr] +
\frac{8\pi \beta k^{3}_{1}k^{2}_{6}(k_{1} - 3k_{4})}{3T^{2}}
- \frac{k_{2}}{2k^{2}_{5} T^{2k_{1} k_{6}}}
\end{equation}
From Equations (\ref{eq61}) and (\ref{eq63}), we observe that the cosmological
constant in both models is a decreasing function of time and it approaches a small 
value as time progresses (i. e. the present epoch), which explains the small value 
of $\Lambda$ at present. \\
{\bf Some Physical Aspects of the Models}: \\
With regard to the kinematic properties of the velocity vector $v^{i}$ in 
the metric (\ref{eq56}), a straight forward calculation leads to the following 
expressions for the scalar of expansion $(\theta)$ and for the shear $(\sigma)$ 
of the fluid.
\begin{equation}
\label{eq64}
\theta = \frac{k^{2}_{1}k_{6}}{T}
\end{equation}
\begin{equation}
\label{eq65}
\sigma^{2} = \frac{k^{2}_{1}k^{2}_{6}(k^{2}_{1} + k^{2}_{4})}{2T^{2}}
\end{equation}
The rotation $\omega$ is identically zero. The non-vanishing components of 
conformal curvature tensor are
\begin{equation}
\label{eq66}
C^{12}_{12} = \frac{k^{2}_{1} k_{6}}{12 T^{2}}\left[k_{1} - 3k_{4} + k_{4}k_{6}
(3k_{1} - k_{4})\right]
\end{equation}
\begin{equation}
\label{eq67}
C^{13}_{13} = \frac{k^{2}_{1} k_{6}}{12 T^{2}}\left[k_{1} + 3k_{4} - k_{4}k_{6}
(3k_{1} + k_{4})\right]
\end{equation}
\begin{equation}
\label{eq68}
C^{14}_{14} = \frac{k^{2}_{1} k_{6}}{6 T^{2}}(k^{2}_{4}k_{6} - k_{1})
\end{equation}
The models represent an expanding, shearing but non-rotating universe in general. 
The models explode with a big bang at $T = 0$ and the expansion in the models stops 
at $T = \infty$. When $k_{1} = 0$ then $\theta = 0$, which implies that $\eta = 0$.
Therefore, viscosity is due to expansion in the model. We take $k_{1} \neq 0$. The
spacetime is Petrov type I non-degenerate. However, if $k_{4} = 0$, the spacetime 
reduces to Petrov type ID. For large values of $T$, the spacetime is conformally flat.
The expressions $\frac{\sigma}{\theta}$ and $\frac{\rho}{\theta^{2}}$ are found to be
\begin{equation}
\label{eq69}
\frac{\sigma}{\theta} = \frac{(k^{2}_{1} + 3k^{2}_{4})^{\frac{1}{2}}}{2\sqrt{3} k_{1}}
\end{equation} 
\begin{equation}
\label{eq70}
\frac{\rho}{\theta^{2}} = \frac{1}{32\pi k^{4}_{1}k^{2}_{6}}\left[k^{2}_{1}k^{2}_{6}
(k^{2}_{1} - k^{2}_{6}) + 4 \Lambda T^{2}\right]
\end{equation}
The rate of expansion $H_{i}$ (Hubble parameters) in the direction of $X$, $Y$, $Z$
are given by
\begin{equation}
\label{eq71}
H_{1} = 0
\end{equation}
\begin{equation}
\label{eq72}
H_{2} = \frac{k_{1}k_{6}(k_{1} + k_{4})}{2T}
\end{equation}
\begin{equation}
\label{eq73}
H_{3} = \frac{k_{1}k_{6}(k_{1} - k_{4})}{2T}
\end{equation}
Since $\lim_{T \rightarrow \infty} \frac{\sigma}{\theta} \neq 0$, hence the models 
do not approach isotropy for large values of $T$. 
\section{Solution in Absence of Shear Viscosity}
When $\eta \to 0$, then the metric (\ref{eq27}) leads to 
\begin{equation}
\label{eq74}
ds^{2} = -16 T^{2} dT^{2} + dX^{2} + T^{(2 + \frac{2L}{M})} dY^{2} + 
T^{(2 - \frac{2L}{M})}dZ^{2}
\end{equation}
The pressure and density for the model (\ref{eq74}) are given by
\begin{equation}
\label{eq75}
8\pi p = \frac{1}{16M^{2}T^{4}}\left[64\pi M^{2} \xi T^{2} + M^{2} 
- L^{2}\right] - \Lambda
\end{equation}
\begin{equation}
\label{eq76}
8\pi \rho = \frac{(M^{2} - L^{2})}{16M^{2}T^{4}} + \Lambda
\end{equation}
\subsection{MODEL I: SOLUTION FOR $\xi = \xi_{0}$}
When $n = 0$, Equation (\ref{eq31}) reduces to $\xi = \xi_{0}$ (constant). Hence 
in this case Equation (\ref{eq75}), with the use of (\ref{eq29}) and (\ref{eq76}), 
leads to
\begin{equation}
\label{eq77}
8\pi (1 + \gamma)\rho = \frac{1}{8M^{2}T^{4}}\left[32\pi M^{2}\xi_{0} T^{2} + M^{2} 
- L^{2}\right]
\end{equation}
Eliminating $\rho(t)$ between (\ref{eq76}) and (\ref{eq77}), we obtain
\begin{equation}
\label{eq78}
(1 + \gamma)\Lambda = \frac{1}{16M^{2}T^{4}}\left[64\pi M^{2}\xi_{0} T^{2} 
- (M^{2} - L^{2})(1 - \gamma)\right]
\end{equation}
\subsection{MODEL II: SOLUTION FOR $\xi = \xi_{0}\rho$}
When $n = 1$, Equation (\ref{eq31}) reduces to $\xi = \xi_{0} \rho$. Hence in 
this case Equation (\ref{eq75}), with the use of (\ref{eq29}) and (\ref{eq76}), 
leads to
\begin{equation}
\label{eq79}
[2(1 + \gamma)T^{2} - \xi_{0}]\rho = \frac{M^{2} - L^{2}}{32\pi M^{2}T^{2}} 
\end{equation}
Eliminating $\rho(t)$ between (\ref{eq76}) and (\ref{eq79}), we have
\begin{equation}
\label{eq80}
[2(1 + \gamma)T^{2} - \xi_{0}]\Lambda = \frac{(M^{2} - L^{2})}{16M^{2}T^{4}}
\left[4T^{2} - 2(1 + \gamma)T^{2} + \xi_{0}\right]
\end{equation}
From Equations (\ref{eq78}) and (\ref{eq80}), we observe that the cosmological
constant in both the models is a decreasing function of time and it approaches a 
small positive value for large time (i. e. the present epoch), when $M^{2} - L^{2}
 > 0$, which is supported by the results from recent type Ia supernovae observations
(Garnavich {\it et al.}, 1998; Perlmutter {\it et al.},1997, 1998, 1999; Riess {\it et al.},
1998; Schmidt {\it et al.}, 1998). \\
{\bf Some Physical Aspects of the Models}: \\
The scalar of expansion$(\theta)$, shear$(\sigma)$ and the non-vanishing components 
of conformal curvature tensor are given by
\begin{equation}
\label{eq81}
\theta = \frac{1}{2 T^{2}}
\end{equation}
\begin{equation}
\label{eq82}
\sigma^{2} = \frac{(M^{2} + 3L^{2})}{48M^{2}T^{4}}
\end{equation}
\begin{equation}
\label{eq83}
C^{12}_{12} = C^{13}_{13} = \frac{(M^{2} - L^{2})}{48 M^{2} T^{4}}
\end{equation}
\begin{equation}
\label{eq84}
C^{14}_{14} = \frac{(L^{2} - M^{2})}{24 M^{2} T^{4}}
\end{equation}
The expressions $\frac{\sigma}{\theta}$ and $\frac{\rho}{\theta^{2}}$ are as follows:
\begin{equation}
\label{eq85}
\frac{\sigma}{\theta} = \frac{1}{2\sqrt{3} M}(M^{2} + 3 L^{2})^{\frac{1}{2}}
\end{equation}
\begin{equation}
\label{eq86}
\frac{\rho}{\theta^{2}} = \frac{1}{32\pi M^{2}}[(M^{2} - L^{2}) + M^{2}T^{4}]
\end{equation}
From Equation (\ref{eq85}), it is observed that the shear is proportional to the
expansion $\theta$ in the models. The models explode with a big bang at $T = 0$
and the expansion in the models stops at $T = \infty$. The spacetime is non-degenerate
Petrov type I in absence of viscosity and if $L = 0$, then spacetime is Petrov type D.
For large values of $T$, the spacetime is conformally flat. Since $\lim_{T \rightarrow 
\infty} \frac{\sigma}{\theta} \neq 0$, hence the models do not approach isotropy for 
large values of $T$. The metric (\ref{eq74}) has singularity at $T = 0$. The models
start from $T = 0$ with big bang and goes on expanding till $T = \infty$. 
\section{Conclusions} 
Some Bianchi type I anisotropic cosmological models with a viscous fluid as the source 
of matter are obtained. Generally, the models are expanding, shearing and 
non-rotating. It is observed that in the presence of viscosity, the spacetime is
Petrov type I D, whereas without viscosity the spacetime is conformally flat. It is
seen that the solutions obtained by Bali and Jain (1997) are particular cases of our 
solutions. In all these models, we observe that they do not approach isotropy 
for large values of time $T$. These models are new and different from those models
obtained by Bali and Jain (1987, 1988) in which free gravitational field was assumed
to be Petrov type D and non-degenerate for Marder (1958).  \\
The cosmological constant in all models given in sections $3.2$ and  $4$ are decreasing 
function of time and they all approach a small positive value as time increases 
(i.e., the present epoch). The values of cosmological ``constant'' for these models 
are found to be small and positive which are supported by the results from recent
supernovae Ia observations recently obtained by the High - Z Supernova Team and 
Supernova Cosmological Project ( Garnavich {\it et al.}, 1998 ; Perlmutter 
{\it et al.}, 1997, 1998, 1999; Riess {\it et al.}, 1998; Schmidt {\it et al.}
, 1998). In section $3.1$, our study is inconsistent with work done by Bali and
Jain (1997). Our study for these models shows constant value of cosmological 
constant for large values of time and do not decrease with time i. e. universe is not
expanding or may be steady state condition. Bali and Jain claim that the universe is 
expanding which does not match with our results. We are trying to find feasible 
interpretation to this case. Further study is in progress.   \\   
\section*{Acknowledgements} 
\noindent
One of the authors (A. Pradhan) thanks to the Inter-University Centre for Astronomy and 
Astrophysics, India for providing  facility under Associateship Programmes where part of 
work was carried out. \\
\newline
\newline

\end{article}
\end{document}